\documentclass{llncs}
\usepackage[utf8]{inputenc}
\usepackage{amsmath,textcomp,amssymb,graphicx,color,textcomp,wasysym}

\usepackage{bbm}

\newcommand{\ZZ}{\mathbbm{Z}}
\newcommand{\NN}{\mathbbm{N}}

\allowdisplaybreaks 
\begin{document} 

\title{Evaluating the quality of local structure approximation using elementary rule 14}
\titlerunning{Elementary rule 14}  
%
\author{Henryk Fuk\'s \and Francis Kwaku Combert}
\authorrunning{Henryk Fuk\'s et al.} 
%

\institute{Department of Mathematics and Statistics, Brock University,\\ St. Catharines,
ON, Canada\\ \email{hfuks@brocku.ca}, \email{fc15uy@brocku.ca}}
\maketitle              

\begin{abstract}
Cellular automata (CA) can be viewed as maps in the space of probability measures. Such maps
are normally infinitely-dimensional, and in order to facilitate investigations of their properties, 
especially in the context of applications, finite-dimensional approximations have been proposed.
The most commonly used one is known as the local structure theory, developed by H. Gutowitz et al. in 1987.
In spite of the popularity of this approximation in CA research, examples of rigorous evaluations of its
accuracy are lacking. In an attempt to fill this gap, we construct a local structure approximation
for rule 14, and study its dynamics in a rigorous fashion, without relying on numerical
experiments. We then compare the outcome with known exact results.

\end{abstract}

\section{Introduction}
One-dimensional elementary cellular automata (CA) can be viewed as maps in the space of probability measures
over bi-infinite binary sequences (to be called \emph{configurations}). This can be understood as follows. Suppose that we start with a large set of  initial
configurations drawn from a certain distribution (for example, from the Bernoulli distribution). 
 Let us now suppose that we apply 
a given  cellular automaton rule to all these configurations. The resulting set of configurations is 
usually no longer described by Bernoulli distribution, but by some other distribution. We can thus say that the CA
rule transforms the initial probability measure into some other measure, and when we apply the local rule again and again, we obtain
a sequence of measures, to be called the \emph{orbit} of the initial measure.

This approach, however,  is not without  difficulties. In order to fully describe a probability measure  over bi-infinite binary sequences, one needs to specify infinitely many \emph{block probabilities}, that is, probabilities of the occurrence of 0, 1, 00, 01, 10, 11, 000, etc -- in short, the probabilities of  occurrence of all possible binary words. This means that the CA rule treated as a map 
in the space of probability measures  is an \emph{infinitely-dimensional map}.

Infinite-dimensional maps are difficult to investigate, even numerically, thus from the early days of CA research, efforts were made
to find a way to approximate them by finite-dimensional maps. In a seminal paper~\cite{gutowitz87a}, published over 30 years ago, 
H. Gutowitz et al. proposed such an approximation, which they called the \emph{local structure theory}. It was an application
 of a well know idea of Bayesian extension, widely used in statistical
physics as a basis of so-called mean-field theories, finite-cluster approximations, and related methods.

Since 1987 the local structure theory has been widely used in CA research, as witnessed by a large number of citations
of  \cite{gutowitz87a}. This could be somewhat surprising, given that relatively few rigorous results are known
about the local structure theory. Usually, the authors using this method simply construct a 
finite-dimensional map or recurrence equations following the recipe given in  \cite{gutowitz87a}, and declare that these posses orbits  approximating the dynamics of the actual CA or related system which they investigate. Judgments on the quality
of the approximation are usually made based on numerical iterations of local structure maps and numerical
simulations of the  CA in question. Numerical results are  thus compared with other numerical results. 

In recent years, however, partial orbits of Bernoulli measures have been computed for some selected elementary CA \cite{paper62}, making 
a somewhat more rigorous approach possible. The goal of this paper is to provide an example of a CA rule for which some block
probabilities are known exactly, and for which local structure equations can be analyzed rigorously, without relying exclusively 
on numerical iterations. This way, the quality of the approximation could be evaluated in a solid and rigorous fashion, without worrying
about numerical errors, finite size effects, etc.

 We selected  elementary CA rule 14 as the most promising example for such study. It has several interesting features: exact probabilities of blocks
of length up to three are known for the orbit of the symmetric Bernoulli measure under this rule, and some of these block probabilities
exhibit non-trivial behaviour - for example, convergence toward the steady state as a power law with fractional exponent.
At the same time, rule 14 conserves the number of pairs 10 \cite{paper23}, and the existence of this additive invariant 
provides a constrain simplifying local structure equations, making them easier to analyze. Since block probabilities of length 3 are known for this rule,
we will construct local approximation  of level~3 and investigate its dynamics not only by simple numerical iterations, but
by finding invariant manifolds at the fixed point and determining the nature of the flow on these manifolds.

One should stress here that in what follows we will use only very minimal formalism. More formal details about the  construction of probability measures over infinite bisequences and the construction of local structure maps for arbitrary rules (both deterministic and
probabilistic) can be found in \cite{paper50}, where the reader will also find more references on these subjects.

\section*{Preliminary remarks about rule 14}
Consider the fully discrete dynamical system (called \emph{cellular automaton}) where $s_i(n)\in\{0,1\}$ is the state of site 
$i\in \ZZ$ at time $n \in \NN$, with dynamics defined by
$
s_i(n+1)=f(s_{i-1}(n), s_{i}(n), s_{i+1}(n)).
$ 
 The function $f: \{0,1\}^3 \to \{0,1\}$ is called the \emph{local rule}. In this paper, we will
consider  $f$ which is defined by
$
 f(x_0, x_1, x_2)= x_1 +x_2 -x_1 x_2-x_0 x_1-x_0 x_2+x_0 x_1 x_2, 
$ 
 and we call the above \emph{rule 14}, following the numbering scheme of Wolfram \cite{Wolfram94}.

Usually, the initial state at $n=0$ is drawn from the Bernoulli distribution, where each site $s_i(0)$ is either
in state 1 with probability $\rho$, or in state 0 with probability $1-\rho$, independently of each other,
where $\rho \in [0,1]$. When $\rho=1/2$, we call this symmetric Bernoulli distribution.

A classical problem in cellular automata theory is to compute the probability of the occurrence of a given binary string 
$\mathbf{a}$
in a configuration obtained  after $n$ iterations of the rule, assuming that the initial configuration
is drawn from the Bernoulli distribution. Such probability will be denoted by $P_n(\mathbf{a})$ and called
 \emph{block probability}.
It is easy to show that if the initial distribution is Bernoulli, then 
the probability of occurrence of $\mathbf{a}$ is independent of its position
in the configuration. We will call such block probabilities \emph{shift invariant}. 

 The set of shift-invariant block probabilities
$P_n(\mathbf{a})$  for all binary strings $\mathbf{a}$ defines a shift-invariant  probability measure on the set of infinite binary bisequences, but we will
not be concerned with the formal construction of such measures here. Interested reader can find all relevant details and references in~\cite{paper50}.

Consider now a configuration in which $s_i(n+1)=1$. By using the definition of rule $f$, one can easily figure
out that $s_i(n+1)$ is determined entirely by the triple $(s_{i-1}(n),s_{i}(n),s_{i+1}(n))$, and that 
the only possible values of $(s_{i-1}(n),s_{i}(n),s_{i+1}(n))$ producing $s_i(n+1)=1$ are $(0,0,1)$,
$(0,1,0)$ or $(0,1,1)$.
This means that probability of obtaining $1$ at time $n+1$ is equal to the sum of probabilities
of ocurrence of blocks $001$, $010$, and $011$ at time $n$, 
$ P_{n+1}(1)=
P_{n}(001)+
P_{n}(010)+
P_{n}(011).
$ 
One can carry out a similar reasoning for longer blocks. For example, a pair of 1s, that is, $s_i(n+1)=1$ and $s_{i+1}(n+1)=1$, can appear only and only if at the previous time step $n$ the lattice positions $i-1,i,i+1,i+2$ 
assumed values 0,0,1,0 or 0,0,1,1, i.e., $(s_{i-1}(n),s_{i}(n),s_{i+1}(n),s_{i+2}(n))=(0,0,1,0)$ or 
$(s_{i-1}(n),s_{i}(n),s_{i+1}(n),s_{i+2}(n))=(0,0,1,1)$. This yields
$
P_{n+1}(11)=
P_{n}(0010)+
P_{n}(0011).
$ 

Obviously, one can write  analogous equations for probabilities of any binary block, obtaining
an infinite system of difference equations.
The complete set of such equations for blocks of length up to 3 for rule 14 is shown below.
\begin{align} \label{probabilitesexact-long}
P_{n+1}(0)&=
P_{n}(000)+
P_{n}(100)+
P_{n}(101)+
P_{n}(110)+
P_{n}(111), \nonumber \\
P_{n+1}(1)&=
P_{n}(001)+
P_{n}(010)+
P_{n}(011), \nonumber \\
P_{n+1}(11)&=
P_{n}(0010)+
P_{n}(0011), \nonumber \\
P_{n+1}(00)&=
P_{n}(0000)+
P_{n}(1000)+
P_{n}(1100)+
P_{n}(1101)+
P_{n}(1110)+
P_{n}(1111), \nonumber \\
P_{n+1}(01)&=
P_{n}(0001)+
P_{n}(1001)+
P_{n}(1010)+
P_{n}(1011), \nonumber \\
P_{n+1}(10)&=
P_{n}(0100)+
P_{n}(0101)+
P_{n}(0110)+
P_{n}(0111) , \nonumber \\
P_{n+1}(000)&=P_n(00000)+P_n(10000)+P_n(11000)+P_n(11100)+P_n(11101)\nonumber \\
&+P_n(11110)+P_n(11111),\nonumber \\
P_{n+1}(001)&=P_n(00001)+P_n(10001)+P_n(11001)+P_n(11010)+P_n(11011),\nonumber \\
P_{n+1}(010)&=P_n(10100)+P_n(10101)+P_n(10110)+P_n(10111),\nonumber \\
P_{n+1}(011)&=P_n(00010)+P_n(00011)+P_n(10010)+P_n(10011),\nonumber \\
P_{n+1}(100)&=P_n(01000)+P_n(01100)+P_n(01101)+P_n(01110)+P_n(01111),\nonumber \\
P_{n+1}(101)&=P_n(01001)+P_n(01010)+P_n(01011),\nonumber \\
P_{n+1}(110)&=P_n(00100)+P_n(00101)+P_n(00110)+P_n(00111),\nonumber \\
P_{n+1}(111)&=0. 
\end{align}
One thing which is immediately obvious is that not all of these equations are independent
because the block probabilities themselves  are not independent. Block probabilities must satisfy
so-called \emph{Kolmogorov consistency conditions}, which are in fact just additivity conditions
satisfied by a measure induced by block probabilities. 
For example, we must have $P_n(1)+P_n(0)=1$, $P_n(01)+P_n(00)=P_n(0)$, etc. Consistency
conditions can be used to express some block probabilities by others. One can show that
for binary strings, among probabilities of blocks of length $k$, only $2^{k-1}$ are independent \cite{paper50},
in the sense that one can choose $2^{k-1}$ block probabilities which are not linked to each other via
consistency conditions. For blocks of length up to $3$, there are 14 block probabilities,
$P_n(0)$,  $P_n(1)$, $P_n(00)$, $P_n(01)$, $P_n(10)$, $P_n(11)$
$P_n(000)$, $P_n(001)$, $P_n(010)$, $P_n(011)$, $P_n(100)$, $P_n(101)$, $P_n(110)$, and $P_n(111)$.
Among them only $2^{3-1}=4$ are independent. While there is some freedom in choosing which ones are
to be treated as independent, we will choose the following four, $P_n(0)$, $P_n(00)$, $P_n(000)$, and
$P_n(010)$. This is called the \emph{short block representation}, and a detailed algorithm for choosing block
this way is described in \cite{paper50}. Here it is sufficient  to say that short block representation ensures that
the blocks selected as independent are the shortest possible ones.

Using consistency conditions, one can now express the remaining blocks of length up to 3 
in terms of $P_n(0)$, $P_n(00)$, $P_n(000)$, and
$P_n(010)$, as follows: 
\begin{align} \label{dependentP}
P_n(1)&=1-P_n(0), \nonumber \\ 
P_n(01)&=P_n(0)-P_n(00), \nonumber \\
P_n(10)&=P_n(0)-P_n(00), \nonumber \\
P_n(11)&=1-2\, P_n(0)+P_n(00), \nonumber \\
P_n(001)&=P_n(00)-P_n(000), \nonumber \\
P_n(011)&=P_n(0)-P_n(00)-P_n(010), \nonumber \\
P_n(100)&=P_n(00)-P_n(000), \nonumber \\
P_n(101)&=P_n(0)-2\, P_n(00)+P_n(000), \nonumber \\
P_n(110)&=P_n(0)-P_n(00)-P_n(010), \nonumber \\
P_n(111)&=1-3\, P_n(0)+2\, P_n(00)+P_n(010).
\end{align}
Using the above substitutions one can reduce eqs. (\ref{probabilitesexact-long}) to the following set of four
equations,
\begin{align}\label{probexact}
                   P_{n+1}(0) &= 1 -P_n(0)  + P_n(000),\\
                 P_{n+1}(00) &= 1 - 2 P_n(0) + P_n(00) + P_n(000), \nonumber \\
   P_{n+1}(000) &= 1  - 3 P_n(0) + 2 P_n(00) + P_n(000) + P_n(010) - P_n(01000), \nonumber \\
                  P_{n+1}(010) &= P_n(0) - 2 P_n(00) + P_n(000). \nonumber 
\end{align}
Note that the above cannot be iterated, because on the right hand side, in addition to the four 
aforementioned independent probabilities, we have probability $P_n(01000)$, the probability of the 
block of length 5.

Fortunately, in spite of the above problem, if the initial Bernoulli measure is symmetric, exact expressions
for probabilities $P_n(0)$, $P_n(00)$, $P_n(000)$ and $P_n(010)$ for rule 14 (that is, the solution of 
eqs. (\ref{probexact}))
can be obtained by combinatorial methods.
We will quote the relevant results below, omitting the proof, which can be found in
\cite{paper34}.
\begin{proposition}[Fuk\'s et al. 2009]
For elementary rule 14, if the initial configuration is drawn from symmetric
Bernoulli distribution, the probabilities of block of length up to 3 are given by
\begin{align} 
 P_n(0)&=\frac{1}{2}\left(1+\frac{2 n-1}{4^n}C_{n-1} \right), \label{P0}\\
 P_n(00)&={2}^{-2-2\,n}(n+1)C_{{n}}+\frac{1}{4},\label{P00}\\
 P_n(000)&= 2^{-2n-3}\left( 4\,n+3 \right) C_{n},\label{P000}\\
 P_n(010)&={2}^{-2-2\,n} \left( n+1 \right) C_{{n}}\label{P010},
\end{align}
where $C_n$ is the $n$-th Catalan number,  $C_n=\frac{1}{n+1} \binom{2n}{n}$.
\end{proposition}
Note that although the above proposition provides  probabilities of $P_n(0)$, $P_n(00)$, $P_n(000)$ and $P_n(010)$
only,  the remaining probabilities of blocks of length up to 3 can be easily computed using eqs. 
(\ref{dependentP}).

Although we know exact solution of eqs. (\ref{probexact}), we can also attempt to obtain an approximate solution
by approximating the ``problematic'' block probability $P_n(01000)$. 
 There exists a method for approximating longer block probabilities by 
probabilities of shorter blocks. This method is called the \emph{Bayesian extension}, and it is known to produce
block probabilities satisfying consistency conditions~\cite{paper50}.
Applying the Bayesian extension to $P_n(01000)$, one obtains
\begin{equation}
 P_n(01000) \approx \frac{P_n(010)P_n(100)P_n(000)}{P_n(10)P_n(00)}.
\end{equation}
In the above, by definition, the
fraction on the right hand side is  considered to be zero whenever its denominator  is equal to zero. Using eqs. (\ref{dependentP}) we can now express
 $P_n(01000)$ in terms of our four independent block probabilities,
\begin{equation}
 P_n(01000) \approx \frac{P_n(010)\left(P_n(00)-P_n(000)\right) P_n(000)}{\left( P_n(0)-P_n(00) \right)P_n(00)}.
\end{equation}
If we replace $P_n(01000)$ in eqs. (\ref{probexact}) by the above approximation, we will
obtain the system of four coupled difference equations,
\begin{align} 
x_{n+1} = &-x_n+z_n+1, \label{eqx} \\
y_{n+1}= &-2\,x_n+y_n+z_n+1, \label{eqy}\\
z_{n+1}= &1+z_n+v_n-3\,x_n+2\,y_n-{\frac {v_n \left( y_n-z_n \right) z_n}{y_n \left(x_n-y_n \right) 
}},\label{eqz}\\
v_{n+1}= &x_n-2\,y_n+z_n, \label{eqv}
\end{align}
where for brevity we introduced variables
$x_n=P_n(0)$, $y_n=P_n(00)$,  $z_n=P_n(000)$. and $v_n=P_n(010)$.
Equations (\ref{eqx})--(\ref{eqv}) will be referred to as \emph{local structure equations} of level 3,
following nomenclature of \cite{gutowitz87a,paper50}.
The designation ``level 3'' pertains to the fact that we used block probabilities of length up to 3.

\section{Exact solutions vs. local structure approximation}
How does the orbit of local structure equations  (\ref{eqx})--(\ref{eqv}) compare with known exact solutions
given by eq. (\ref{P0})--(\ref{P010})? 
In order to find this out, we will assume that the initial probability measure is symmetric  Bernoulli,
meaning that $x_0=P_0(0)=1/2$, $y_0=P_0(00)=1/4$, $z_0=P_0(000)=1/8$, and $v_0=P_0(010)=1/8$.
\begin{figure}
\begin{center}
\includegraphics[width=5in]{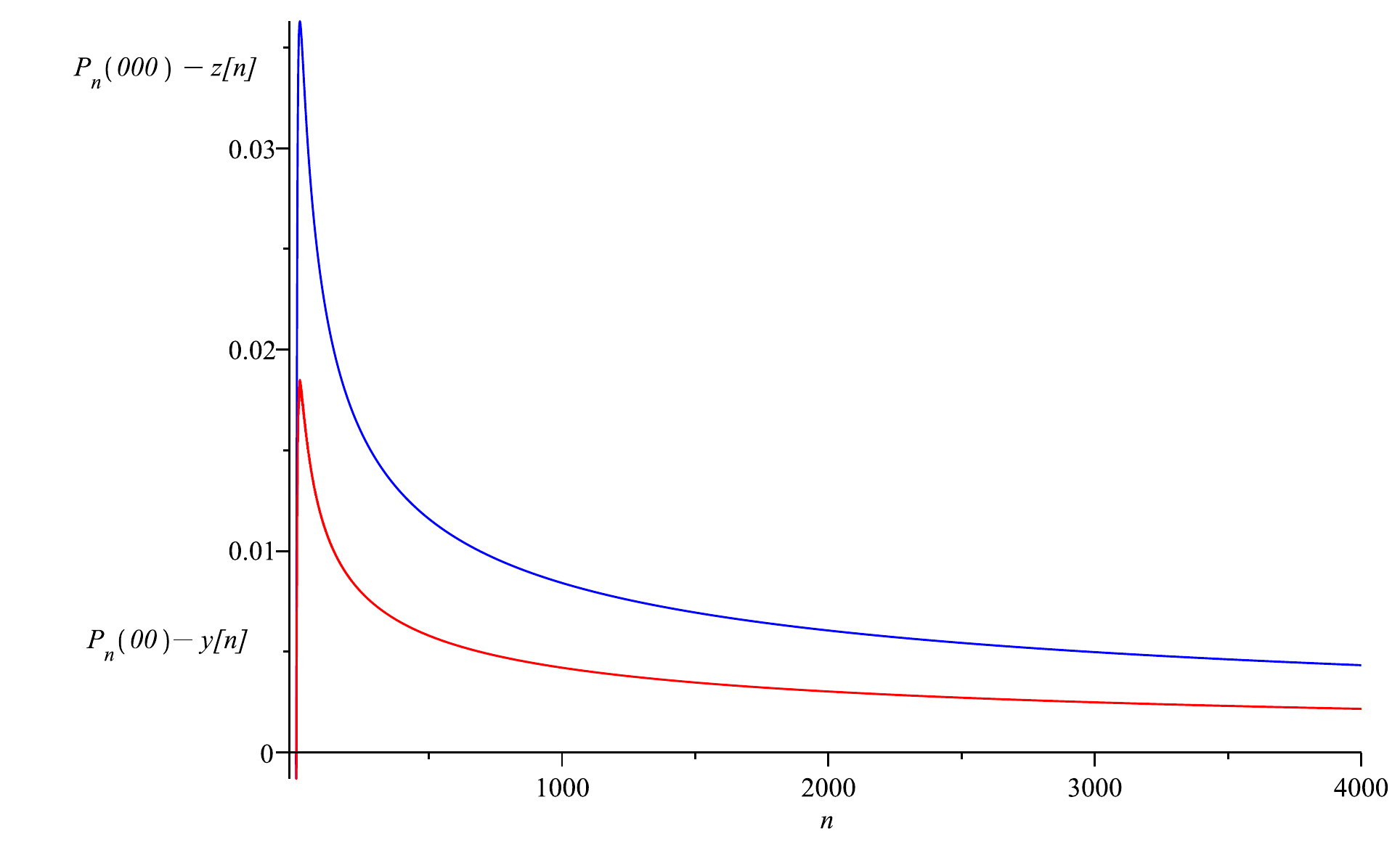}
\end{center}
\caption{Differences between exact and approximate values of block probabilities as a function of $n$. Two differences are shown,  $P_n(00)-y_n$ (lower curve) and $P_n(000)-z_n$ (upper curve).}\label{exactvslst}
\end{figure}
Figure \ref{exactvslst} shows differences between exact probabilities obtained from eq. (\ref{P0})--(\ref{P010})
and values obtained by iterating local structure equations  (\ref{eqx})--(\ref{eqv}). Two differences are shown,  $P_n(00)-y_n$ (lower curve) and $P_n(000)-z_n$ (upper curve). In both cases we can see that the 
difference tends to zero as $n\to \infty$. Values of $P_n(0)-x_n$  and $P_n(010)-v_n$ (not shown)
exhibit similar behaviour.

This indicates that even though the local structure approximation of level 3 does not produce exact
values of block probabilities at finite $n$, it seems to become exact in the limit $n \to \infty$.
To verify this, let us first note that from eq. (\ref{P0})--(\ref{P010}) we obtain
 \begin{align*}
 \lim_{n \to \infty} P_n(0)=\frac{1}{2}, \quad
 \lim_{n \to \infty} P_n(00)=\frac{1}{4},\quad
 \lim_{n \to \infty} P_n(000)= 0, \quad
 \lim_{n \to \infty} P_n(010)= 0.
 \end{align*}
 We will denote these values by $(x^{\star},y^\star,z^\star,v^\star)=(\frac{1}{2}, \frac{1}{4}, 0, 0)$.
One can easily verify that
 $(x^{\star},y^\star,z^\star,v^\star)$ is a fixed point of eqs.  (\ref{eqx})--(\ref{eqv}).
In what follows, we will investigate stability of this fixed point. We will prove that
the following property holds.
 \begin{proposition}\label{4dconvergence}
    If the dynamical system given by eqs.  (\ref{eqx})--(\ref{eqv})  is iterated starting from
initial conditions  $x_0=1/2$, $y_0=1/4$, $z_0=1/8$, and $v_0=1/8$, then 
 $\lim_{n \to \infty} (x_{n},y_{n}, z_{n}, v_{n}) = (x^\star,y^\star,z^\star,v^\star)=\left(\frac{1}{2}, \frac{1}{4}, 0, 0\right)$.
 \end{proposition}
This means that the local structure map approximates the exact probabilities remarkably well,
converging to the same fixed point as the exact values. We will prove Proposition~\ref{4dconvergence}
by reducing local structure equations to two dimensions and by computing local manifolds at the fixed point.

\section*{Reduction to two dimensions}
Close examination of equations (\ref{eqx})--(\ref{eqv}) reveals some obvious
symmetries. First of all, it is easy to check that $x_{n+1}-y_{n+1}=x_n-y_n$. Since $x_0-y_0=\frac{1}{4}$, we have
$x_n-y_n=\frac{1}{4}$ for all $n$, thus
\begin{equation}\label{xreplace}
 x_n=y_n+\frac{1}{4}.
\end{equation}

Further simplification is possible. Note that 
$v_{n+1}-y_{n+1}=3(x_n-y_n)-1=3 \cdot \frac{1}{4}-1=-\frac{1}{4}$. 
This implies that for any $n>0$,
$ v_{n+1}=y_{n+1}-\frac{1}{4}$, 
or, equivalently, that for any $n>1$,
\begin{equation} \label{vreplace}
 v_{n}=y_n-\frac{1}{4}.
\end{equation}
Note that this does not hold for $n=0$, because in this case $v_0=y_0-1/8$.
Now, using eqs. (\ref{xreplace}) and (\ref{vreplace}), we can reduce our dynamical system to two dimension,
as eqs. (\ref{eqy}) and (\ref{eqz}) become
\begin{align*}
y_{n+1}= &-2\,(y_n+\frac{1}{4})+y_n+z_n+1, \\
z_{n+1}= &1+z_n+(y_n-\frac{1}{4})-3\,(y_n+\frac{1}{4})+2\,y_n-{\frac {(y_n-\frac{1}{4}) \left( y_n-z_n \right) z_n}{y_n \left((y_n+\frac{1}{4})-y_n \right) 
}}.
\end{align*}
After simplification we obtain, for $n\geq1$,
\begin{align}
y_{n+1}= &\frac{1}{2}-y_n+z_n, \label{redy} \\
z_{n+1}= &z_n-{\frac {(4y_n-1) \left( y_n-z_n \right) z_n}{y_n}} \label{redz},
\end{align}
where we start the recursion at $n=1$, taking $y_1=3/8$, $z_1=7/32$. The last two values were obtained by 
direct computation of $y_1$ and $z_1$ from eqs. (\ref{eqy}) and (\ref{eqz}) for $n=0$, by substituting
$x_0=1/2$, $y_0=\frac{1}{4}$, $z_0=v_0=1/8$ on the right hand side.

We will prove the following result.
\begin{proposition}\label{2dconvergence}
   If the dynamical system described by eqs. (\ref{redy}) and (\ref{redz}) is iterated starting
at $y_1=3/8$, $z_1=7/32$, then
 \begin{equation}\label{conject1}
  \lim_{n \to \infty} (y_{n}, z_{n}) = \left(\frac{1}{4},0\right).
\end{equation}
\end{proposition}
In order to prove the above proposition let us first denote
$   \mathbf{x} = \begin{bmatrix}
           y \\
           z \\           
         \end{bmatrix}$. 
In this notation, eqs. (\ref{redy}) and (\ref{redz}) define  two-dimensional map
\begin{align}\label{differ1}
\mathbf{F}(\mathbf{x})=
     \begin{bmatrix}
           \frac{1}{2}-y+z  \\
           z-{\frac {(4y-1) \left( y-z \right) z}{y}}            
         \end{bmatrix}.
\end{align}
It is easy to check that the map $\mathbf{F}$ has the fixed point  $\mathbf{x}^{\star}= \left[ \begin{array}{l}
\frac{1}{4}\\0
\end{array}\right]$. In order to prove Proposition \ref{2dconvergence}, all we need is to show that 
$\mathbf{x}^{\star}$ is asymptotically stable (or at least semi-stable in the relevant domain).

The Jacobian matrix of $\mathbf{F}$ evaluated at the fixed point $\mathbf{x}^{\star}$ is given by
\[ A=\left[ \begin{array}{ccc}
-1 & 1 \\
0 & 1  \end{array} \right], 
\] 
and its eigenvalues are $-1$ and $1$. Since these eigenvalues have an absolute value equal to 1, 
 the fixed point $\mathbf{x}^{*}$ is a non-hyperbolic fixed point and one cannot determine its stability by eigenvalues
alone. We will investigate its stability by resorting to the center manifold theory.

Let $P$ be the matrix of column eigenvectors of $A$, and let $P^{-1}$ be its inverse, 
\[ P=\left[ \begin{array}{rr}
1 & \frac{1}{2} \\
0 & 1  \end{array} \right],\\\qquad P^{-1}=\left[ \begin{array}{rr}
1 & -\frac{1}{2} \\
 0 &  \text{ } 1  \end{array} \right].\]
We will first move the fixed point to the origin and simultaneously diagonalize the linear part of $\mathbf{F}$.
The following change of variables accomplishes this task,
\begin{align} \label{transxX}
 \mathbf{X}= P^{-1} (\mathbf{x} - \mathbf{x}^{\star}), 
\end{align}
where the components of the new variable $\mathbf{X}$ will be denoted by $Y$ and $Z$. Eq. (\ref{transxX}) thus yields 
\begin{align}
 Y&=y-\frac{1}{4}-\frac{z}{2},\\
 Z&=z.
\end{align}
Change of variables from $\mathbf{x}$ to $\mathbf{X}$ transforms the dynamical system $\mathbf{x}_{n+1}=\mathbf{F}(\mathbf{x}_n)$
into the system
\begin{align}
\mathbf{X}_{n+1}=P^{-1}\mathbf{F}(P\mathbf{X}_n+\mathbf{x}^{\star})-P^{-1}\mathbf{x}^{\star}.
\end{align}
This yields, after simplification,
\begin{align}
Y_{n+1} &= -Y_{n}+\frac{1}{2}\frac{\left(4Y_{n}+2Z_{n}\right)\left(Y_{n}-\frac{1}{2}Z_{n}+\frac{1}{4}\right)Z_{n}}{Y_{n}+\frac{1}{2}Z_{n}+\frac{1}{4}}, \label{diagY}\\
Z_{n+1}&= Z_{n}-\frac{\left(4Y_{n}+2Z_{n}\right)\left(Y_{n}-\frac{1}{2}Z_{n}+\frac{1}{4}\right)Z_{n}}{Y_{n}+\frac{1}{2}Z_{n}+\frac{1}{4}}.\label{diagZ}
\end{align}
One can immediately see that the above system has $(0,0)$ as a fixed point, and that its linear part
is  given by
$Y_{n+1} = -Y_{n}$, 
$Z_{n+1}= Z_{n}$.
As mentioned earlier, there is nothing we can say about the stability of $(0,0)$ by examining the linear part alone,
except that in the vicinity of $(0,0)$ the $Y$ variable is changing its sign at each iteration. We will use the method
outlined in~\cite{Perko2008} to find the invariant manifold corresponding to $-1$ eigenvalue. We will call this manifold
the \emph{flip manifold} and denote it by $W^f$.

Let us assume that $W^f$ has the equation $Z=h(Y)$, where $h$ in the vicinity of $0$  is given by 
the series $h(Y)=a_2{Y}^{2}+a_{{3}} {Y}^{3}+a_{{4}}{Y}^{4}+a_{{5}} {Y}^{5}+ \ldots$. 
Note that the series starts from the quadratic term, and this is because the manifold $Z=h(Y)$ must be tangent to
the $Y$ axis (we already diagonalized our dynamical system).

The condition for invariance of $W^f$ requires that the relationship $Z_n=h(Y_n)$ remains valid in the next time step,  meaning that $Z_{n+1}=h(Y_{n+1})$. 
Let us rewrite eqs. (\ref{diagY}) and (\ref{diagZ}) as
\begin{align}
Y_{n+1} &= G_1(Y_n,Z_n),\\
Z_{n+1}&= G_2(Y_n,Z_n),
\end{align}
where
\begin{align}
G_1(Y,Z)&= -Y+\frac{1}{2}\frac{\left(4Y+2Z\right)\left(Y-\frac{1}{2}Z+\frac{1}{4}\right)Z}{Y+\frac{1}{2}Z+\frac{1}{4}}, \\
G_2(Y,Z)&= Z-\frac{\left(4Y+2Z\right)\left(Y-\frac{1}{2}Z+\frac{1}{4}\right)Z}{Y+\frac{1}{2}Z+\frac{1}{4}}.
\end{align}
Condition $Z_{n+1}=h(Y_{n+1})$ now becomes 
$
 G_2(Y,Z)=h(G_1(Y,Z)),
$
and, by taking $Z=h(Y)$, it yields
\begin{equation}
 G_2(Y,h(Y))=h(G_1(Y,h(Y))).
\end{equation}
This means that if we expand  $G_2(Y,h(Y))-h(G_1(Y,h(Y)))$ into the Taylor series with respect to $Y$, all
coefficient of the expansion should be zero. Such expansion, done by the Maple symbolic algebra system, yields
\begin{gather}\label{taylorse}
G_2(Y,h(Y))-h(G_1(Y,h(Y)))=\left( 2\,a_{{3}}-4\,a_{{2}} \right) {Y}^{3}+ \nonumber \\
 \left( -4\,a_{{3}}-4
\, \left( 4+\frac{1}{2}a_{{2}} \right) a_{{2}}+16\,a_{{2}}+4\,{a_{{2}}}^{2}
 \right) {Y}^{4}+          
\nonumber \\ 
 \bigg( 2\,a_{{5}}-4\,a_{{4}}-4\, \left( 4+\frac{1}{2}a_{{
2}} \right) a_{{3}}-8\,a_{{2}}a_{{3}}+16\,{a_{{2}}}^{2}+16\,a_{{3}} 
\nonumber \\
- \left( -4\,a_{{3}}-4\, \left( 4+\frac{1}{2}a_{{2}} \right) a_{{2}}+16\,a_{{
2}} \right) a_{{2}} \bigg)     
{Y}^{5}+O \left( {Y}^{6} \right).  \nonumber
\end{gather}
Coefficients in front of $Y^3,Y^4, Y^5, \ldots$ must be zero, yielding the system of equations for
 $a_2, a_3, a_4, \ldots$, 
\begin{align}
0&=2\,a_{{3}}-4\,a_{{2}},\nonumber\\
0&=-4\,a_{{3}}-4
\, \left( 4+\frac{1}{2}a_{{2}} \right) a_{{2}}+16\,a_{{2}}+4\,{a_{{2}}}^{2},      
\nonumber \\ 
0&= 2\,a_{{5}}-4\,a_{{4}}-4\, \left( 4+\frac{1}{2}a_{{
2}} \right) a_{{3}}-8\,a_{{2}}a_{{3}}+16\,{a_{{2}}}^{2}+16\,a_{{3}}\\&-
 \left( -4\,a_{{3}}-4\, \left( 4+\frac{1}{2}a_{{2}} \right) a_{{2}}+16\,a_{{
2}} \right) a_{{2}}, 
\nonumber\\
\ldots \nonumber
\end{align}
Solving the above system one obtains
$ a_{2}= 4$, $a_{3}= 8$, $a_{4}= 3$, $a_{5}= -32$, etc.
The flip manifold $W^f$ is, therefore, given by
\begin{equation}
 Z=h(Y)=4Y^2+8Y^3+32Y^4-32Y^5+O(Y^6).
\end{equation}
By substituting $Z_n$ by $h(Y_n)$ on the right hand side of eq. (\ref{diagY}) and Taylor expanding again one obtains the
equation describing the dynamics on the flip manifold $W^f$,
\begin{align}\label{flipmanifolddynamics}
Y_{n+1}=-Y_{n}+8\,{Y}^{3}_{n}+32\,{Y}^{4}_{n}+O \left( {Y}^{5}_{n} \right). 
\end{align}
The above equation has $0$ as a fixed point, and we need to determine its stability. 
Recall that a fixed point $\bar{x}$ of $x_{n+1}=f(x)$ is said to be \emph{asymptotically stable} if there exist
$\delta>0$ such that for any $x_0$ satisfying $|x_0 -\bar{x}|< \delta$ we have $\lim_{n \to \infty}x_n=\bar{x}$.
We will use
the following general test for asymptotic stability~\cite{Murakami2005}.
\begin{theorem}[Murakami 2005] \label{theorem1}
Let $\bar{x}$  be a fixed point of $x_{n+1}=f(x_n)$. Suppose that $f \in C^{2k-1}(\mathbb{R})$,  $f^{\prime}(\bar{x})=-1$, $f^{j}(\bar{x})=0$ for $j\in\{2,3,\dotsc,k-1\}$, and that $f^{(k)}(\bar{x})\neq 0.$ 
 If $k$ is odd and $f^{(k)}(\bar{x}) > 0$, then $\bar{x} $ is asymptotically stable.
%
%
\end{theorem}
In our case, for eq. (\ref{flipmanifolddynamics}), $f(x)=-x+8\,{x}^{3}+32\,{x}^{4}+O \left( {x}^{5} \right)$, $\bar{x}=0$, $f^\prime(\bar{x})=-1$, $f^{\prime\prime}(\bar{x})=0$, and
 $f^{(3)}(\bar{x})=48$, thus the theorem applies, meaning that zero is  asymptotically stable fixed point of
eq. (\ref{flipmanifolddynamics}).
\begin{figure}[h!]
\begin{center}
 \includegraphics[width=4.0in]{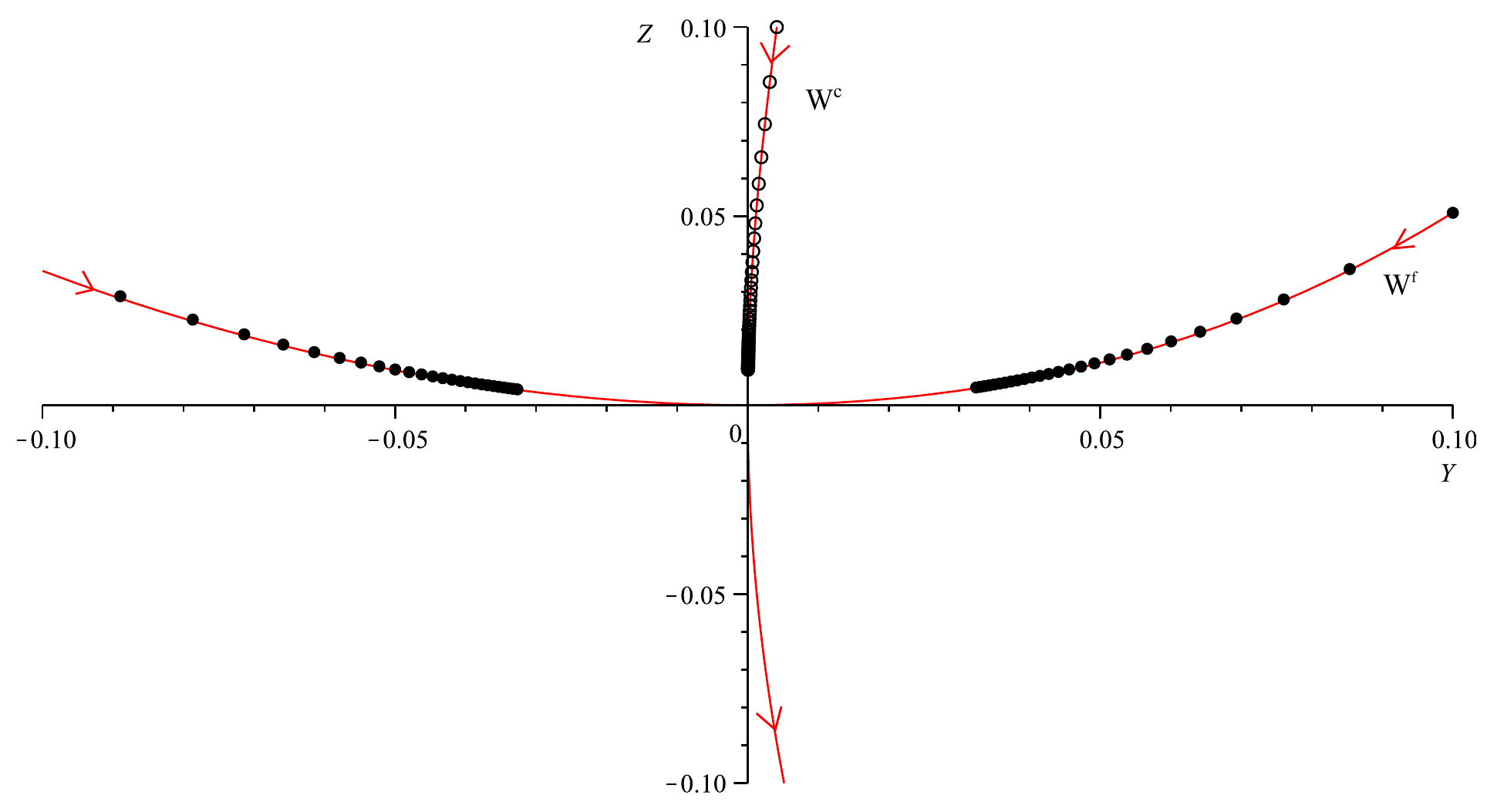}
\includegraphics[width=4.0in]{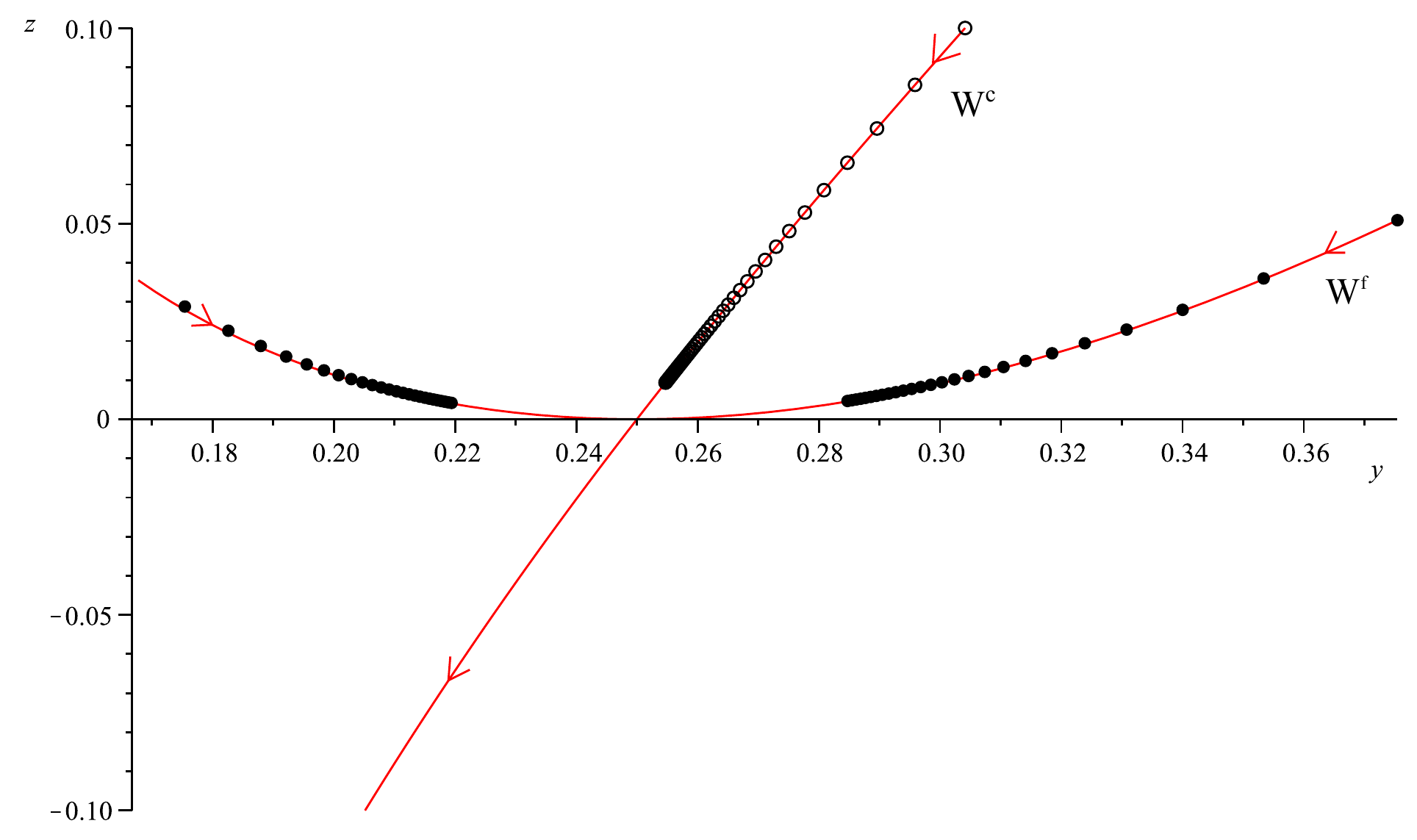}
\end{center}
\caption{The flip manifold $W^f$ and the center manifold $W^c$ in transformed $(Y,Z)$ coordinates  (top) and original $(y,z)$ coordinates
(bottom). Points represent numerically computed orbits of a sample point on $W^c$ (\textopenbullet) and $W^f$ (\textbullet).}\label{manifoldsfig}
\end{figure}

We need to perform a similar analysis for the eigenvalue 1 and the corresponding center manifold $W^c$.
Let us assume that $W^c$ has equation $Y=g(Z)$, where $g$ in the vicinity of $0$  is given by 
the series $g(Z)=b_2{Z}^{2}+b_{{3}} {Z}^{3}+b_{{4}}{Z}^{4}+b_{{5}} {Z}^{5}+ \ldots$. 
The condition for invariance of $W^c$ requires that $Y_n=g(Z_n)$ remains valid at the next time step,  $Y_{n+1}=g(Z_{n+1})$. Using our previous notation this means that
$ G_1(Y,Z)=g(G_2(Y, Z))$, 
which, by substituting $Y=g(Z)$, yields
\begin{equation}
 G_1(g(Z),Z)=g(G_2(g(Z), Z)).
\end{equation}
As before, by expanding $G_1(g(Z),Z)=g(G_2(g(Z), Z))$ into the Taylor series and setting all
coefficient of the expansion to  be zero we obtain, using Maple, 
$b_2 = \frac{1}{2}$, $b_3 = -\frac{1}{2}$, $b_4 = -4, b_5 = -\frac{3}{2}$, etc. 
The equation of the center manifold is,  therefore,
\begin{equation}
Y= \frac{1}{2} Z^2-\frac{1}{2} Z^3 -4 Z^4 -\frac{3}{2} Z^5 + O\left(Z^6\right).
\end{equation}
By substituting $Y_n$ by $g(Z_n)$ on the right hand side of eq. (\ref{diagZ}) and Taylor expanding again one obtains the
equation describing the dynamics on the center manifold $W^c$,
\begin{align}\label{centermanifolddynamics}
Z_{n+1}= Z_n-2 Z_n^2+6 Z_n^3 -6 Z_n^4 + O \left( Z_n^{5} \right). 
\end{align}

In order to determine the stability of 0 in the above difference equation, let us first define semistability.
A fixed point $\bar{x}$ of $x_{n+1}=f(x)$ is said to be \emph{asymptotically semistable from the right} if there exist
$\delta>0$ such that for any $x_0$ satisfying $x_0 -\bar{x}< \delta$ we have $\lim_{n \to \infty}x_n=\bar{x}$.
One can show \cite{Elyadi1999} that if $f^\prime(\bar{x})=1$ and $f^{\prime \prime}(\bar{x})<0$ then $\bar{x}$ is
assymptotically stable from the right. In our case, for eq. (\ref{centermanifolddynamics}), we have 
$f(x)=x-2 x^2+6 x^3 -6 x^4 + O \left( x^{5} \right)$, $\bar{x}=0$, 
$f^\prime(\bar{x})=1$ and  $f^{\prime \prime}(\bar{x})=-4<0$, thus for eq. (\ref{centermanifolddynamics}), zero is asymptotically semistable from the right.

Figure \ref{manifoldsfig} shows manifolds $W^f$ and $W^c$ together with sample orbits generated numerically
by iterating eqs.  (\ref{diagY}) and (\ref{diagZ}). Direction of the flow is indicated by arrows. Note that 
$W^c$ is asymptotically semistable only on the right (for $Z>0$), and unstable on the left (for $Z<0$).
The left-sided instability is irrelevant for us, since $Z$ represents the probability of 000 block, thus
it must always be positive.

Since 0 is asymptotically stable on $W^f$, and asymptotically semistable on $W^c$, we conclude that for $Z_0>0$, $\lim_{n \to \infty} (X_n, Z_n)=(0,0)$, or, equivalently,
$\lim_{n \to \infty} (x_n, z_n)=(1/4,0)$,
as claimed in Proposition~\ref{2dconvergence}. Proposition~\ref{4dconvergence} follows automatically. $\Square$

\section{Quality of local structure approximation}
We have demonstrated so far that for rule 14, the local structure approximation of level 3 reproduces
correctly the limiting values of probabilities of blocks  of length up to 3. What about the \emph{rate}
of convergence to these limiting values?
In order to find this out, let us consider rates of convergence to zero of  $P_n(000)$ and its approximation $z_n$.
We know that 
$
  P_n(000)=2^{-2n-3}\left( 4\,n+3 \right) C_{n}$, 
where
$
C_n=\frac{1}{n+1} \binom{2n}{n}=\frac{(2n)!}{n!(n+1)!}.
$ 
Using Stirling's formula for large $n$, 
$\displaystyle n!\sim {\sqrt {2\pi n}}\left({\frac {n}{e}}\right)^{n},
$  
the Catalan number $C_n$ can be approximated as
$$C_n
=\frac{1}{n+1}\frac{(2n)!}{{(n!)}^2}
\sim \frac{1}{n+1}\frac{{\sqrt {4\pi n}}\left({\frac {2n}{e}}\right)^{2n} }   {{\left( {\sqrt {2\pi n}}\left({\frac {n}{e}}\right)^{n}  \right)}^2}
= \frac{1}{n+1} \frac{2^{2n}}{\sqrt{\pi n }} = \frac{1}{n+1} \frac{4^{n}}{\sqrt{\pi n }},
$$
meaning that $P_n(000)$ converges toward zero as a power law $P_n(000)\varpropto n^{-1/2}$,
where $x \varpropto y$ means the ratio $x/y$ tends  to a positive number as $n\to \infty$.

\begin{figure}
\begin{center}
\includegraphics[width=4.0in]{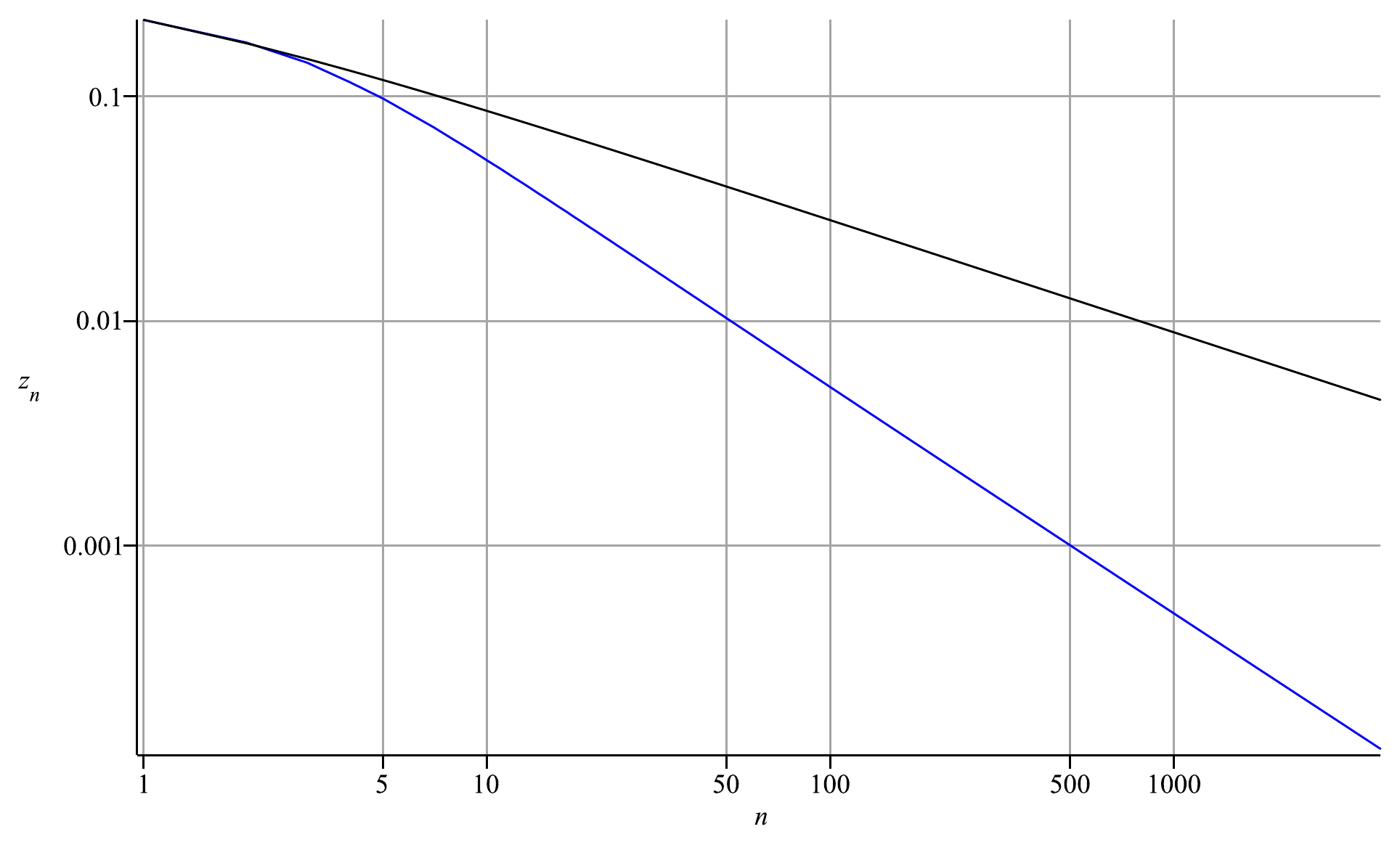}
\end{center}
\caption{Plot of $P_n(000)$ (upper line) and its local structure approximation $z_n$ (lower line) as a function of $n$
in log-log coordinates.}\label{powerlawv}
\end{figure}

Let us now examine convergence of $z_n$ to 0.
We do not have a formula for $z_n$, but we can generate $z_n$ numerically, by iterating the local structure equations.
Figure~\ref{powerlawv} shows the graph of $z_n$ vs. $n$ in log-log coordinates together with the graph
of $P_n(000)$ vs. $n$. We can see that both graphs appear to be almost straight lines, confirming that
both $z_n$ and $P_n(000)$ behave as $n^\alpha$ for large $n$. The difference is in the value of the exponent $\alpha$.
For $P_n(000)$ the exponent (computed as a slope of the upper line in Figure~\ref{powerlawv}) is $\alpha\approx-1/2$,
whereas for $z_n$ the exponent  (computed as a slope of the lower line) is $\alpha\approx-1$. 

The value of the exponent  $\alpha\approx-1$ can be explained as follows. The starting point
of the local structure approximation orbit, $y_1=3/8$, $z_1=7/32$, lies almost on the center manifold $W_c$. The 
convergence toward the fixed point is, therefore, dominated by eq. (\ref{centermanifolddynamics}),
which, if we keep only leading terms, becomes $Z_n=Z_n-2Z_n^2$.
Although this equation is not solvable in a closed form, we can 
obtain  its asymptotic solution using the standard technique used in the theory of iterations of complex analytic functions.
We can namely conjugate the map $Z \to Z -2Z^2$ with appropriate M\"{o}bius
transformation, which moves the fixed point to $\infty$  \cite{Beardon91,Devaney89}.
In our case, the M\"{o}bius map will simply be the inverse, meaning that we change variables
in the equation $Z_n=Z_n-2Z_n^2$ to $u_n=1/Z_n$, obtaining
\begin{equation}
 u_{n+1}=u_n+2 + \frac{4}{u_n-2}.
\end{equation}
Since $u_n \to \infty$, the above can be approximated for large $n$  by $u_{n+1}= u_n +2$, 
which has the solution $u_n =  2t+u_0$, or, going back to the original variable, $Z_n = \frac{1}{2t+1/z_0}$. The result $z_n=Z_n \varpropto t^{-1}$ immediately follows.

In conclusion, one could thus say
that the local structure approximation  correctly reproduces not only the coordinates of the the fixed point but also the \emph{type} of convergence toward
the fixed point (as a power law). It fails, however, to reproduce the correct value of the exponent in the power law.
This in agreement with the commonly reported results of investigations of critical phenomena: mean-field type
theories cannot reproduce values of fractional exponents in power laws.


It would be interesting and beneficial to extend results of this paper to non-symmetric initial Bernoulli measures.
Numerical evidence suggests that local structure approximation remains exact in the limit of $n \to \infty$ in
such cases, but to be sure one would need to generalize eqs. (\ref{P0})--(\ref{P010}) to non-symmetric initial
measure. This, in principle, should be possible, and will be attempted in the future.\\
\begin{scriptsize}\noindent\textbf{Acknowledgement:} H.F. acknowledges financial support from the Natural Sciences and
Engineering Research Council of Canada (NSERC) in the form of Discovery Grant.                                                                              \end{scriptsize}

\end{document}